# Topological Insulator $V_x Bi_{1.08-x}Sn_{0.02}Sb_{0.9}Te_2S$ as a Promising *n*-type Thermoelectric Material

Lei Chen[a,b], Weiyao Zhao[a,b], Meng Li[a], Guangsai Yang[a,b], Lei Guo[c], Abudulhakim Bake[a], Peng Liu[a,b], David Cortie[a,b], Ren-Kui Zheng[d], Zhenxiang Cheng[a], Xiaolin Wang[a,b*]

[a] Institute for Superconducting and Electronic Materials, Australian Institute for Innovative Materials, University of Wollongong, North Wollongong, 2500, Australia

[b] ARC Centre of Excellence for Future Low-Energy Electronics Technologies (FLEET), University of Wollongong, North Wollongong, NSW 2522, Australia

[c] School of Physics, Southeast University, Nanjing 211189, China

[d] School of Physics and Materials Science, Guangzhou University, Guangzhou 510006, China

**Abstract**

As one of the most important *n*-type thermoelectric (TE) materials, $Bi_2Te_3$ has been studied for decades, with efforts to enhance the thermoelectric performance based on element doping, band engineering, etc. In this study, we report a novel bulk-insulating topological material system as a replacement for *n*-type $Bi_2Te_3$ materials: V doped $Bi_{1.08}Sn_{0.02}Sb_{0.9}Te_2S$ (V:BSSTS) . The V:BSSTS is a bulk insulator with robust metallic topological surface states. Furthermore, the bulk band gap can be tuned by the doping level of V, which is verified by magnetotransport measurements. Large linear magnetoresistance is observed in all samples. Excellent thermoelectric performance is obtained in the V:BSSTS samples, e.g., the highest figure of merit ZT of ~ 0.8 is achieved in the 2% V doped sample (denoted as $V_{0.02}$) at 550 K. The high thermoelectric performance of V:BSSTS can be attributed to two synergistic effects: (1) the low conductive secondary phases $Sb_2S_3$, and $V_2S_3$ are believed to be important scattering centers for phonons, leading to lower lattice thermal conductivity; and (2) the electrical conductivity is increased due to the high-mobility topological surface states at the boundaries. In addition, by replacing one third of costly tellurium with abundant, low-cost, and less-toxic sulfur element, the newly produced BSSTS material is inexpensive but still has comparable TE performance to the traditional $Bi_2Te_3$-based materials, which offers a cheaper plan for the electronics and thermoelectric industries. Our results demonstrate that topological materials with unique band structures can provide a new platform in the search for new high performance TE materials.

**Introduction**

Thermoelectric materials can harvest ambient waste heat to produce high-quality electricity, thus have attracted increasing interest from both academia and industries as an essential role in green energy technologies[1-3]. The heat-to-electricity efficiency is governed by the figure-of-merit of thermoelectric materials, defined as $ZT = S^2\sigma T/\kappa$, where $S$ is Seebeck coefficient, $\sigma$ is electronic conductivity, $T$ is absolute temperature, and $\kappa$ is thermal conductivity taking up the contributions from electrons $\kappa_e$ and lattice vibrations $\kappa_L$[4, 5]. It is intuitive that good thermoelectric materials should favor a combined large $S$ and $\sigma$, and small $\kappa$. However, some perverse trade-offs prohibiting an infinite $ZT$, for instance, $S$ and $\sigma$ are usually competing on carrier concentration, while $\kappa_e$ is proportional to $\sigma$ via Wiedemann-Franz law[6]. Such that the decoupling of these terminologies is highly desired, which has been realized in several phonon-glass electron-crystal (PGEC) skutterudites[7-9] and phonon-liquid electron-crystal (PLEC) superionic materials[10-12], with their decoupled electrical and thermal transport rendering ultrahigh $ZT$s. Recently, topology conception has been radically introduced to thermoelectric research, where the topological insulators (TIs), featured by insulating bulk states and time-reversal-symmetry-protected conducting boundary states, are an emerging type of thermoelectric materials with totally decoupled electron and phonon pathways[13-17].

There is a rich history of screening feasible thermoelectric materials from vast TIs, and some famous thermoelectric materials, such as $Bi_2Se_3$, $Bi_2Te_3$, $Sb_2Te_3$, and SnTe are indeed discovered in this way[16-24]. A theoretical prediction claims that the gapless edge states mainly contribute to anomalous Seebeck effect; the $ZT$ is no more intrinsic but being dependent on the geometric size of the Dirac band, which provides another degree-of-freedom to be manipulated[15]. The TIs enjoy several advantages as thermoelectric materials, including: i) sizable Seebeck effect with respect to scattering time arising from the boundary-bulk interactions; ii) appropriate carrier concentration from narrow band gap due to usually containing heavy elements with large electronegativity, and iii) high carrier mobility (yet small effective mass) of the boundary Dirac band. With some specific presumptions satisfied, e.g. optimized geometric size or Fermi level, the $ZT$ of TIs is directly correlated to their microscale size parameter, which can be facilely tuned to enhance $ZT$ significantly greater than 1[15].

Quintuple layered $Bi_2Te_3$ is one of the most investigated bulk TIs with outstanding room-temperature thermoelectric performance in both *p*-type and *n*-type regimes[25-28]. On the basis of crystallography, its outermost two Te layers are Van de Wall bonded with each other and covalently bonded with Bi atoms, and the inner Te layer is octahedrally bonded in lower energy; while two intercalated Bi layers are energetically equivalent. It has been reported *ZT* of ~1.8 in *p*-type $Bi_2Te_3$-$Sb_2Te_3$ binary alloys[29], with Sb atoms randomly substituting Bi sites. By contrast, the counterpart *n*-type thermoelectric $Bi_2Te_3$ are still in niche application, where extraordinary efforts have been paid to $Bi_2Te_3$-$Bi_2Se_3$ binary alloys with Se atoms preferably substituting Te sites in higher energy[30-32]. However, the *ZT* is generally limited below unit even with further doping, not to mention other existing issues, including the small intrinsic electron effective mass in this system thus poor Seebeck coefficient, and the low abundance of Bi (0.048 ppm) and high toxicity of Se elements[30-32]. It is proved that by using sulfur to replace one third of tellurium in $Bi_2Te_3$, the band gap can be enlarged and the bipolar effect is suppressed, so the materials can be applied at higher temperature[33, 34]. Rather than randomly distributing in the Te position, the S intends to form a S layer by totally occupying the position of $-Te^{(2)}-$, which may also contribute to phonon scattering[34].

The newly discovered *n*-type TI, (V, Sn) co-doped $Bi_{1.1}Sb_{0.9}Te_2S$ (V:BSSTS), has inspired us as a potential alternative for the $Bi_2Te_3$-$Sb_2Te_3$ alloys for near-room-temperature thermoelectric application[13]. V:BSSTS has outstanding topological insulating properties with tenable bulk band gap of ~0.34 eV and preserved quantum oscillations to above 50 K, but its thermoelectric properties are still ambiguous. In this work, for the first time, we studied the thermoelectric properties of polycrystalline V:BSSTS at near room temperature. We found the crystal structure and phase distribution of V:BSSTS being analogous to $Bi_2Te_3$ based on various characterization techniques. When sintered into highly dense thermoelectric pellets, V:BSSTS show a promising *ZT* of ~0.6 at 300 K, plateaued to ~0.8 at 550 K, which are explained by the correspondent thermoelectric and transport measurements from 3 K to 550 K. Our work clarifies that V:BSSTS a feasible *n*-type room-temperature thermoelectric material with less toxicity, lower cost, and comparable performance to the widely-used $Bi_2Te_3$-based alloys.

**Experimental details**

**Synthesis:** High-purity reagent V, Sn, Bi, Sb, Te, and S in trace metal basis (3N, Sigma-Aldrich Australia) were mixed with stoichiometric ratio and sealed in evacuated silica tubes. The mixture were heated to 1100 °C in 10 h, soaked at this temperature for 24 h, then naturally cooled to room temperature. The obtained ingots were pulverized into fine powder by planetary ball milling, which was then loaded into a graphite mould in 10 mm diameter and consolidated by spark plasma sintering (SPS) at 350 °C for 5 min under a uniaxial pressure of 50 MPa. The obtained pellets were shaped into round disks and rectangular bars for thermal and electrical measurements, respectively.

**Characterization:** The phase purity of synthesized samples was identified by powder X-ray diffraction (XRD) on a GBC MMA system equipped with graphite monochromatized Cu Kα radiation ($\lambda$=1.5418 Å). The microstructure and phase distribution were studied on field-emission scanning electron microscope (SEM, JEOL JSM-7500) by secondary electron (SE), backscattered electron (BSE) and energy-dispersive X-ray spectroscopy (EDS) analysis.

**Thermoelectric and Transport Property Measurement:** The Seebeck coefficient ($S$) and electrical conductivity ($\sigma$) were measured using a four-probe method in vacuum (commercial RZ2001i). The thermal conductivity was calculated from $\kappa = D \times C_p \times \rho$, where $D$, $C_p$ and $\rho$ are thermal diffusion, heat capacity and density, respectively. Here, $D$ was measured by a laser flash method (LFA 1000, LINSEIS) in vacuum, $C_p$ was measured using a differential scanning calorimetry (DSC-204F1, NETZSCH) in Argon atmosphere, $\rho$ was measured by Archimedes' method.

The low-temperature transport property, namely $\sigma$ between 3-380 K was measured by magnetoresistance and the Hall effect using a Quantum Design Physical Properties Measurement System (PPMS, 14 T, Quantum Design). The five-contact Hall-bar geometry was adopted. The carrier concentration ($n$) and mobility ($\mu$) were calculated from the formula $n = 1 / (e \times R_H)$ and $\mu = \sigma \times R_H$, where $e$ is electron charge, and $R_H$ is Hall coefficient.

**Results and Discussion**

Previous report on V:BSSTS has revealed that V can tune the bulk band gap to make it dramatically more insulating, e.g., the band gap of single crystals varies from 0.11 eV ($V_{0.02}$) to 0.34 eV ($V_{0.08}$).[35] As shown in Fig. 1(a), the bulk band gap is positively related to V doping level, referring to a lower bulk carrier concentration, while the surface states are of the same magnitude among the three samples. It is reasonable to consider that the bulk and surface states both contribute to the thermoelectric properties. Fig. 2 shows the room-temperature XRD patterns of the $V_{0.00}$, $V_{0.02}$ and $V_{0.08}$ samples. All diffraction peaks can be well indexed to the hexagonal $Sb_2SeTe_2$ (JCPDS card, 26-0659) except one weak impurity peak, indicating that the main tetradymite phase is obtained. As predicted, the nanostructure of topological insulator is a key factor to good thermoelectric performance. Therefore, to achieve good thermoelectric properties, we employ ball milling to prepare fine powder samples before SPS, building more grain boundaries. As schematically illustrated in Fig. 1(b), the grain boundaries possess better conductivity than the bulk due to the topological surface states, while scattering the phonons to reduce the thermal conductivity.

Fig. 3(a), (b) and (c) show SEM (BSE mode) images and EDS element mapping images of the $V_{0.00}$, $V_{0.02}$, and $V_{0.08}$ samples, respectively. From these images it can be seen that low-conductivity, micrometer-size secondary phases exist in all the samples, and increase with the doping level of V. By analyzing the element distribution via EDS, which are shown in Fig. 3(d)-(f) one can deduce that the low conductive secondary phases are $Sb_2S_3$, and $V_2S_3$. These secondary phases are believed to be important scattering centers for phonons, thus leading to lower lattice thermal conductivity[36-38].

Fig. 4(a) shows the temperature-dependent electrical conductivity for the V:BSSTS system with three different vanadium doping levels from 3K to 550 K. Although *n*-type $Bi_2Te_3$ and the V:BSSTS system are both topological insulators, they have a significant difference: the $\sigma$-$T$ curve of $Bi_2Te_3$ indicates metal-like behavior, while V:BSSTS shows semiconductor-like behavior. Note that the Fermi level of *n*-type $Bi_2Te_3$ penetrates into the bulk conduction band, but in V:BSSTS, the Fermi level is still in the bulk band gap, which is responsible for the large Seebeck coefficients of these samples. Therefore, compared with metallic topological insulators, the bulk-insulating topological insulator is a better system to study the surface-enhanced thermoelectric behavior. Let us focus on the $V_{0.00}$ sample, in which the total conductivity decreases monotonically from 550 K to ~30 K, suggesting that the insulating bulk carriers dominate the transport behavior in this temperature region. With further cooling, the conductivity shows a plateau below 30 K, meaning that the metallic surface carriers contribute to the σ-T behavior. In the V doped samples, the bulk-dominant conductivity behavior is much weaker: 1) in the $V_{0.02}$ sample, the high temperature conductivity shows less temperature dependence, and the low temperature (< 30 K) conductivity shows a sign of an upturn; 2) in the $V_{0.08}$ sample, with a band gap of ~ 0.34 eV, as reported by a single crystal study[35], the conductivity below ~150 K shows an upturn, indicating that the surface carriers contribute more to the total transport property than in the undoped and low doped samples. The phenomenon that we found in polycrystalline samples agrees well with previous single crystal results[35], suggesting that the good topological insulating properties, e.g., wide bulk gap, high surface state survival temperature, exist in the polycrystalline thermoelectric samples fabricated via SPS.

Furthermore, Hall effect and magnetoresistance ($MR = \frac{R(B)-R(0)}{R(0)}$) measurements were also conducted on the V:BSSTS samples with magnetic field up to 14 T perpendicular to the current, and the results are shown in Fig. 5(a)-(c), respectively. Based on the Hall effect results, the carrier concentrations and mobilities were obtained and are presented in Fig. 4(b). The carrier densities of all three samples are between $3 \times 10^{18}$ and $1.5 \times 10^{19}$ cm$^{-3}$, with weak temperature dependence. The total carrier density decreases with V doping, which shows similar behavior to the σ-T curves, and is sensitive to the bulk band gap. Moreover, the mobility of the carriers in three samples at ~380 K is between 250 and 400 cm$^2$/Vs, which is close to the reported $Bi_2Te_3$ data.[39] In the $V_{0.00}$ sample, the MR curve shows a linear feature (especially in the low-field region) without any sign of saturation, reaching a value of ~13% at 3 K and 14 T. On heating up, the maximum MR value decreases monotonically to ~5.2% at 300 K and 14 T. The low-field linear feature of the MR curves remains up to ~ 100 K, above which, it shows parabolic-like behavior. In the strong field region, however, e.g., > 3 T, the MR curves are have a linear appearance, even at 300 K. In the V doped samples, the 14 T MR values are higher than for the $V_{0.00}$ sample, e.g., ~ 27% for the $V_{0.02}$ sample at 3 K and 14 T, and ~37% for the $V_{0.08}$ sample at 3 K and 14 T. The linear-like MR features, generally and at high temperatures only in strong fields, are observed also in the V doped samples. It should be noted that the MR behavior in the polycrystalline samples made by SPS is quite similar to that in single crystals, but with relatively smaller MR values. Similar room-temperature linear-MR behavior was also reported for $Bi_2Te_3$ nanosheets, in which, at 300 and 340 K, the MR curves show low-field-parabolic-like and strong-field-linear-like behavior, due to the linear dispersion of the band structure. In our case, the band structure is quite similar to that reported for $Bi_2Te_3$ nanosheets, while the material itself (as shown in Fig. 3(b)) can be compared with high-density, mosaic-like, nanoscale TI sheets. The MR behavior is a good evidence that the topological surface states exist in our V:BSSTS, at least at room temperature.

Fig. 6(a) shows the electrical conductivity curves from room temperature to 550 K, which has been discussed in connection with Fig. 4(a) together with the low temperature conductivity. Fig. 6(b) shows the temperature dependence of the Seebeck coefficient for our V:BSSTS samples. The negative Seebeck coefficient values indicate that our samples are *n*-type, in which electrons are the major carriers. The Seebeck coefficient values of the $V_{0.00}$ and $V_{0.02}$ samples increase at first and then decrease with increasing temperature. This is because from 300 K to about 425 K, the increasing Seebeck coefficient is mainly attributable to the enhanced phonon-electron interaction with increasing temperature. As the temperature further increases, the valence band and the conduction band will experience thermal excitation of the charge carriers, thereby generating electrons in the conduction band and holes in the valence band, and the Seebeck coefficient then decreases due to the bipolar effects. Interestingly, the Seebeck values of the $V_{0.08}$ sample do not change much when temperature is below 380 K, but above that, it decreases significantly. According to a previous report on the V:BSSTS single crystals, both *n*-type bulk carriers and *p*-type surface carriers can coexist. In the undoped and low V-doped single crystals $V_{0.00}$ and $V_{0.02}$, the bulk carriers are dominant, so they are *n*-type, while the highly V-doped single crystal $V_{0.08}$ shows *p*-type behavior due to the significant contribution of its surface carriers[35]. In our $V_{0.08}$ polycrystalline sample, however, the bulk carriers are still dominant due to the high defect level. The *n*-type bulk carriers and *p*-type surface carriers contribute to different Seebeck effects[40] and therefore, to unconventional Seebeck behavior. The power factors calculated from $PF = S^2\sigma$ are shown in Fig. 6(c), in which it can be seen that the $V_{0.08}$ sample has the lowest power factor as a result of low electrical conductivity and a low Seebeck coefficient. The $V_{0.02}$ sample possesses the highest PF among all of the three samples over the whole temperature range due to its proper bulk band gap, as well as its topological surface contributions.

Fig. 6(d) illustrates the thermal conductivity of the three samples as a function of temperature. It can be seen from that the thermal conductivity is effectively reduced by a higher vanadium doping level. In order to obtain a deeper insight into the thermal transport behavior, the lattice and electronic parts of the total thermal conductivity were calculated, as shown in Fig. 6(e). Based on the Wiedemann-Franz law[41], the electronic thermal conductivity ($\kappa_e$) is determined by the electrical conductivity via $\kappa_e = L\sigma T$, where $L$ is the Lorenz number. The Lorenz number for the $Bi_2Te_3$ system was calculated using a single parabolic band (SPB) model. The lattice thermal conductivity ($\kappa_L$) then can be calculated by $\kappa_L = \kappa - \kappa_e$. As shown in the inset of Fig. 6(e), the electronic contribution ($\kappa_e$) to the total thermal conductivity of the V:BSSTS samples increases with increasing temperature. On the other hand, $\kappa_L$ obviously decreases with increasing doping level, due to the enhanced phonon scattering by the increasing number of secondary phases and defects.

The calculated $ZT$ values of the three samples are presented in Fig. 6(f). The temperature-dependent behavior of $ZT$ of the two doped samples show that $ZT$ first decreases with increasing temperature, and then increases, finally achieving a maximum ZT of ~0.8 for $V_{0.02}$ at ~450 K. The $ZT$ in the undoped sample shows similar behavior to the doped samples when the temperature is below 450 K, and the highest $ZT$ value of ~0.6 is obtained at 450 K. The $ZT$ goes down again when heated above 450 K, and the value of $ZT$ was reduced to ~0.5 at ~550 K. This was because the thermal conductivity of the undoped sample kept increasing at high temperature, while those of the two doped samples remained flat. The $ZT$ behavior of the undoped, low vanadium doped, and high vanadium doped samples revealed that, compared with the undoped sample, $V_{0.02}$ sample presents the most appropriate band gap and strong surface states, resulting in the best thermoelectric behavior in the V:BSSTS system.

Besides the high peak $ZT$ value, a high average $ZT_{ave}$ of 0.68 for the $V_{0.02}$ sample over the whole measurement temperature range (297 K – 550 K) was also obtained. The $ZT_{ave}$ values were calculated based on the relationship

$$ZT_{ave} = \frac{1}{T_h-T_c}\int_{T_c}^{T_h} ZT dT \tag{1}$$

where $T_h$ and $T_c$ are the hot-side and cold-side temperatures, respectively. Thus, a high $ZT_{ave}$ value is relevant to thermoelectric device fabrication and practical applications. This is

because the thermal-electric conversion efficiency $\eta$ of a thermoelectric power generator is given by

$$\eta = \frac{T_h - T_c}{T_h} \frac{\sqrt{1 + (ZT)_{ave}} - 1}{\sqrt{1 + (ZT)_{ave}} + T_c/T_h}. \quad (2)$$

## Conclusion

We synthesized V:BSSTS samples by the melting-cooling method, and then employed SPS to make condensed pellets of samples with different V doping levels: 0.00, 0.02, and 0.08, based on the band structure from a previous single crystal study[35]. A comprehensive magnetotransport study was carried out on the polycrystalline samples, which suggested that resistivity curves of the SPS samples share similar temperature and magnetic responses. Specifically, bulk-insulating behavior was observed in the $\sigma$-$T$ curves, as evidenced by decreasing conductivity with cooling, which are, more or less, interrupted by the surface contributions, e.g., the conductivity shows a plateau in $V_{0.00}$ sample below 30 K, an upturn below ~30 K in the $V_{0.02}$ sample, and an upturn below ~150 K in the $V_{0.08}$ sample. With a perpendicular magnetic field, the magnetoresistance shows a roughly linear increase in all the samples, even at room temperature, which is strong evidence that the linearly dispersed topological surface states contribute to the transport behavior. We thus measured the surface-state-enhanced thermoelectric behavior, and found: 1) among the three samples, the $V_{0.02}$ sample presents the best thermoelectric performance; 2) the maximum Seeback coefficient of 172 $\mu$V/K was obtained in the $V_{0.02}$ sample at ~450 K, together with the largest power factor of ~9.5 $\mu$W/K$^2$cm; 3) the thermal conductivity of $V_{0.02}$ sample is close to that of the non-V-doped sample, but is decreased in the $V_{0.08}$ sample, showing the lowest thermal conductivity of ~ 0.4 W/m K. The best $ZT$ in this system is ~0.8, which was found in the $V_{0.02}$ sample at ~450 K, along with $ZT_{ave}$ of 0.68 in the temperature range of 297 K – 550 K. We obtained excellent thermoelectric performance in a sulfur substituted $n$-Bi$_2$Te$_3$-based bulk-insulating topological insulator material, which offers another avenue to achieve high performance with cheaper materials. Moreover, the bulk-insulating system also provides a good platform to explore the dynamics in surface-state-enhanced thermoelectric performance.

# Acknowledgements:

This work was partially supported by the Australian Research Council (ARC) through an ARC Professorial Future Fellowship project (FT 130100778, XLW) and a Linkage Infrastructure Equipment and Facilities (LIEF) Grant (LE120100069, XLW).

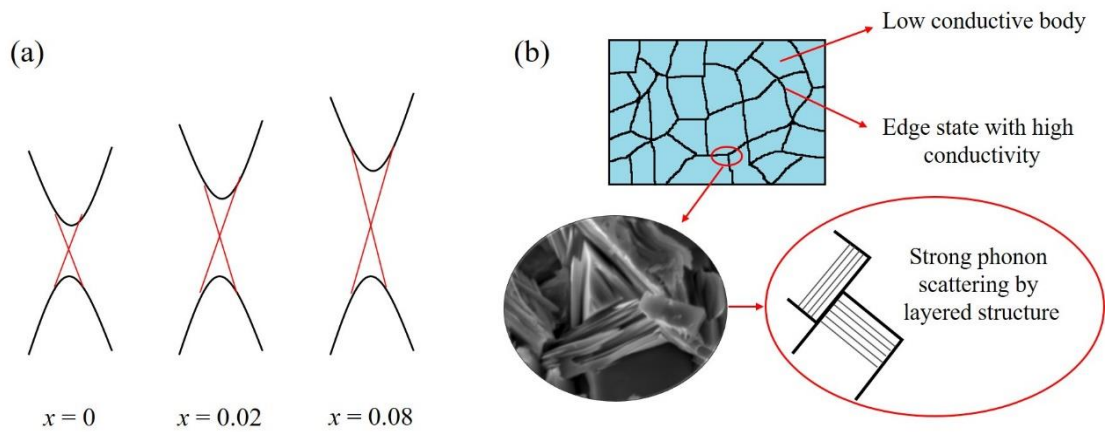

Fig. 1 (a) Diagram of the band structures of V:BSSTS with different V doping. (b) Schematic illustration of how electronic current and heat flow through the grain boundaries.

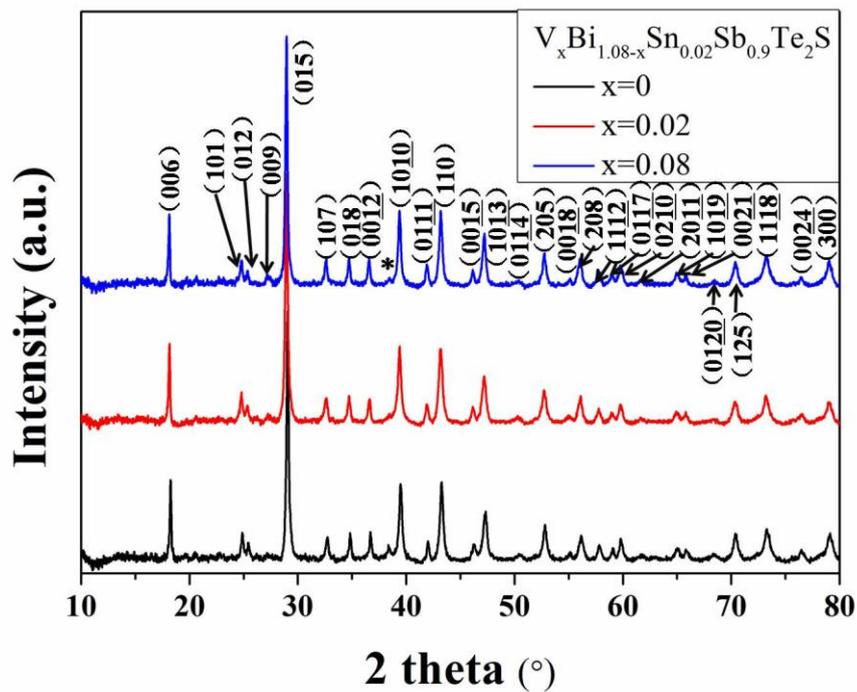

Fig. 2 XRD patterns of $V_{0.00}$, $V_{0.02}$ and $V_{0.08}$ samples. * is the impurity peak.

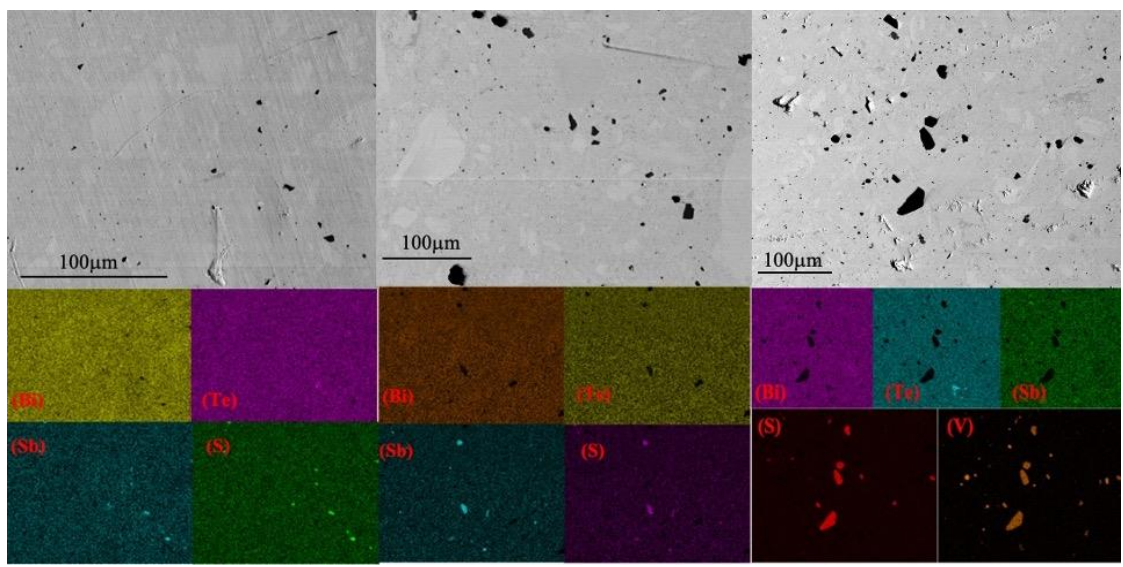

Fig. 3 (a), (b), (c) SEM (BSE) images of the $V_{0.00}$, $V_{0.02}$, and $V_{0.08}$ samples, respectively. (d), (e), (f) EDS mapping of (a), (b) and (c), respectively.

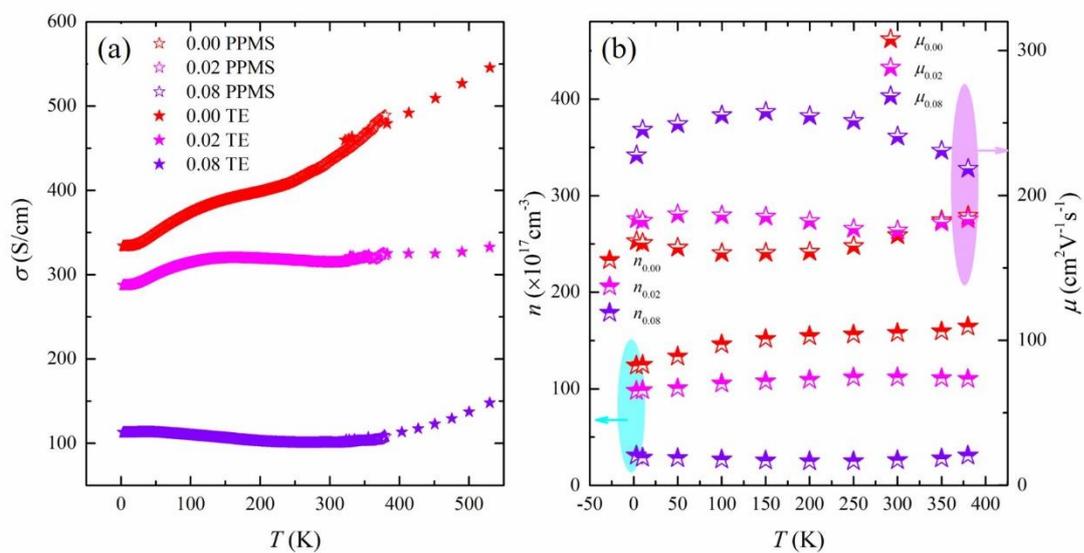

Fig. 4 Electronic transport properties of the V:BSSTS samples. (a) The temperature dependence of the conductivity for the three samples, measured by PPMS (3-380 K) and thermoelectric equipment (330-550 K). (b) The carrier density and mobility between 3 and 380 K, obtained from the Hall effect measurements.

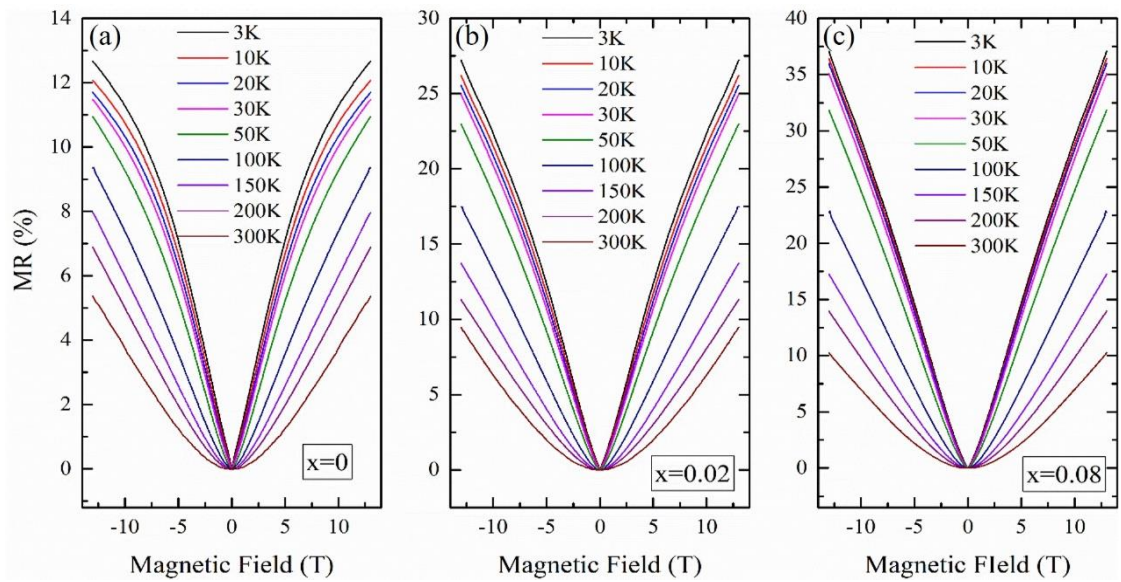

Fig. 5 Magnetoresistance of V:BSSTS samples at various temperatures from 3 to 300 K.

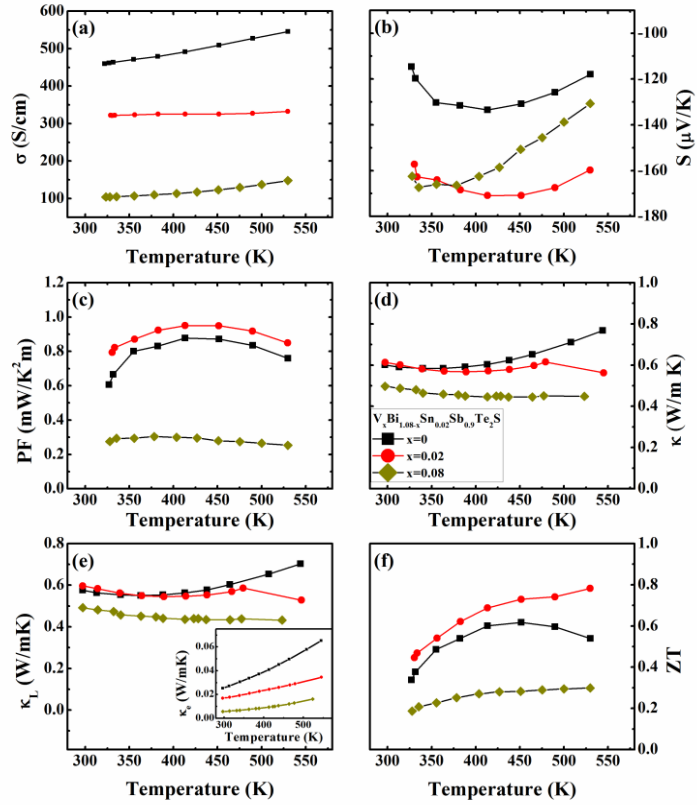

Fig. 6 Temperature-dependence of the thermoelectric properties of the V:BSSTS samples: (a) electrical conductivity; (b) Seebeck coefficient; (c) power factor; (d) thermal conductivity; (e) lattice thermal conductivity $\kappa_L$, with inset showing the electronic thermal conductivity $\kappa_e$; and (f) the figure of merit *ZT*.